\documentclass[conference, transmag, 10pt]{IEEEtran}

\usepackage{amsmath,amssymb,amsfonts}
\usepackage[style=numeric, backend=bibtex, backref=false, sorting=none]{biblatex} 
\usepackage{algorithmic}
\usepackage{graphicx}
\usepackage{textcomp}

\usepackage{float}
\usepackage{csquotes}

\usepackage{flushend}

\usepackage{tikz}
\usepackage{lipsum}

\usepackage{tabularx}
\usepackage{array}

\usepackage{caption}

\usepackage{letltxmacro}
\LetLtxMacro\oldttfamily\ttfamily
\DeclareRobustCommand{\ttfamily}{\oldttfamily\csname ttsize\endcsname}
\newcommand{\setttsize}[1]{\def\ttsize{#1}}

\setttsize{\small}

\usepackage{xcolor}

\def\BibTeX{{\rm B\kern-.05em{\sc i\kern-.025em b}\kern-.08em
    T\kern-.1667em\lower.7ex\hbox{E}\kern-.125emX}}

\DeclareSourcemap{
	\maps{
		\map{ % Replaces '{\_}', '{_}' or '\_' with just '_'
			\step[fieldsource=url,
			match=\regexp{\{\\\_\}|\{\_\}|\\\_},
			replace=\regexp{\_}]
		}
		\map{ % Replaces '{'$\sim$'}', '$\sim$' or '{~}' with just '~'
			\step[fieldsource=url,
			match=\regexp{\{\$\\sim\$\}|\{\~\}|\$\\sim\$},
			replace=\regexp{\~}]
		}
		\map{ % Replaces &
			\step[fieldsource=url,
			match=\regexp{\{\\\x{26}\}},
			replace=\regexp{\x{26}}]
		}
		\map{ % Replaces #
			\step[fieldsource=url,
			match=\regexp{\{\#\}|\{\\#\}|\{\\\#\}},
			replace=\regexp{\#}]
		}
	}
}

\usepackage[hidelinks]{hyperref}

\bibliography{bibliography}

\begin{document}

\title{A Practical Multilevel Governance Framework for \\ Autonomous and Intelligent Systems}
\thanks{}

\author{\IEEEauthorblockN{Lukas D. Pöhler}
\IEEEauthorblockA{\textit{Dept. Computer Engineering} \\
	\textit{Technical University of Munich}\\
Munich, Germany\\
lukas.poehler@tum.de}
\and
\IEEEauthorblockN{Klaus Diepold}
\IEEEauthorblockA{\textit{Dept. Computer Engineering} \\
	\textit{Technical University of Munich}\\
	Munich, Germany\\
	kldi@tum.de}
\and
\IEEEauthorblockN{{Wendell Wallach} }
\IEEEauthorblockA{\textit{Technology and Ethics Research Group} \\
	\textit{Yale Interdisciplinary Center for Bioethics} \\
	New Haven, USA \\
	wendell.wallach@yale.edu}
}

	\maketitle
	\thispagestyle{plain}
	\pagestyle{plain}

	\begin{abstract}
	Autonomous and intelligent systems (AIS) facilitate a wide range of beneficial applications across a variety of different domains. However, technical characteristics such as unpredictability and lack of transparency, as well as potential unintended consequences, pose considerable challenges to the current governance infrastructure. Furthermore, the speed of development and deployment of applications outpaces the ability of existing governance institutions to put in place effective ethical-legal oversight. New approaches for agile, distributed and multilevel governance are needed. This work presents a practical framework for multilevel governance of AIS. The framework enables mapping actors onto six levels of decision-making including the international, national and organizational levels. Furthermore, it offers the ability to identify and evolve existing tools or create new tools for guiding the behavior of actors within the levels. Governance mechanisms enable actors to shape and enforce regulations and other tools, which when complemented with good practices contribute to effective and comprehensive governance.
	\end{abstract}

% ----- %%%%%%%%%%%%%%%%%%%%%%%%%%%%%%%%%%%%%%%%%%%%%%%%%%%%%%%%%%%%% ----- % 
% ----	SECTION Introduction
% ----- %%%%%%%%%%%%%%%%%%%%%%%%%%%%%%%%%%%%%%%%%%%%%%%%%%%%%%%%%%%%% ----- %

\section{Introduction}
\label{S_Intro}

\subsection{AIS bring Chances and Challenges}

With recent advances in computing and data availability, powerful autonomous and intelligent systems (AIS) and their broad deployment have become possible in a wide array of domains and at a large scale. Beneficial applications of AIS exist in different fields ranging from the medical sector including improved diagnosis, treatment and possibilities for tailored and affordable prostheses to the energy sector with efficient energy production, distribution and usage. AIS allow advanced analytics, prediction, and risk mitigation for environmental and food security applications. The recent availability of large language models (LLMs) and other generative AI frameworks has placed power tools in the hands of the public, and is transforming education. Given such vast potentials, AIS practically contribute to all 17 UN SDGs~\cite{WEF2018, Chu2018good}.

However, challenges arise with the availability of AIS technologies and their deployment. From a societal perspective, AIS might be deployed in ways that erode trust between individuals and decrease political stability. Examples for this are human-like online agents such as social bots that propagate false information and are potentially able to sway public opinion, unintended online echo chambers that reinforce biased opinions, automated and personalized disinformation, and influence campaigns for both marketing and political purposes~\cite{Brundage2018}. An important factor here is the rapid progress of generative models in deep learning for video~\cite{Kim2018Deep}, text~\cite{Radford2019LanguageMA} and audio generation~\cite{Gao2018} that can be easily distributed and whose output is increasingly harder for humans to detect. Recent advances are even of concern to researchers as demonstrated by the piece-wise release of the language model GPT-2 as well as the widely distributed ChatGPT (GPT3.5) in November 2022. From an arms control perspective, AIS pose several challenges for national and international security and stability~\cite{POEHLER2018GGE}. For example, AIS convergence with life sciences might result  in the creation of harmful organisms by a robotized laboratory~\cite{Brockmann2019}. AIS convergence with cyber-technology can lead to automated cyber-attacks~\cite{Brundage2018}. AIS also challenge international stability with lethal autonomous weapons systems that might deploy nuclear weapons or other high-powered munitions~\cite{Sipri2019}. Scholars increasingly expect that the nature of war will change significantly with a high degree of autonomy in weapons systems~\cite{Singer2009, Cummings2017, Rickly2017, Scharre2018, Gill2019}. From a technical risk perspective, AIS could pose unintended consequences that were not anticipated during development. These might result from the technical immaturity of systems, careless engineering process, corrupt or biased data. Risk mitigation strategies for AIS are challenging. Due to the complex nature and self-modifying characteristics of machine-learning components the assessment of risks is extremely difficult~\cite{Pringle2017}.

\subsection{The Changing Governance Landscape}

Considering these challenges a central question is how to effectively govern AIS development and deployment in order to harness their benefits while mitigating potential negative consequences or undesired societal impacts. Conventional top-down governance systems have increasingly faced difficulties in coping with new and emerging technologies. This is often the result of diverging multi-stakeholder interests or the self-interest of actors with little interest in collaboration. In addition, traditional governance approaches are challenged by the rapid development and wide-scale deployment of technologies whose scope for  impact is broad.  Scholars identify this governance challenge as a pacing problem, which describes a gap between slow regulatory processes based upon static views of technology, and  rapidly deployed disruptive applications~\cite{Marchant2011}. 

This is particularly the case for AIS, a powerful suite of technologies which incorporate components that are often developed in a distributed and collaborative way as open-source tools. Furthermore, digital technologies have a potentially high impact on society and international stability because their rapid deployment often encompasses and shapes whole societies. What further challenges the governance of AIS is differing ways a specific technology can be applied in a variety of domains. For example, algorithms on autonomous path planning can be incorporated in medical robots, self-driving vehicles or military systems. Therefore, simply focusing on the development and engineering stage of an AIS is not sufficient. A more holistic approach to governance including the deployment of an AIS in each domain and across its lifecycle is necessary to prevent undesired societal impacts. This responsibility for comprehensive oversight goes beyond governments and includes companies, industry initiatives, research institutes, civil society groups, affected individuals and other stakeholders.

\subsection{The Need for a Multilevel Governance Framework}

Because of the complexity of AIS and the multitude of relevant actors, several governance approaches in addition to traditional governmental regulation and regulatory bodies are needed. These approaches apply to various decision-making levels ranging from companies with ethical guidelines or codes of conduct, to professional organizations with moral standing that set and design principles and standards. This fragmentation of the governance landscape makes it challenging to keep track of the relevant actors with existing governance tools. However, for effective governance of AIS, actors must be both aware of governance tools and ensure that their application is addressed or enforced when necessary. The 2019 report of the UN Secretary-General’s High-level Panel on Digital Cooperation identifies six gaps in current arrangements of digital cooperation: low focus on digital technology and cooperation on policy agendas, non-inclusive standard-setting organizations, overlap among mechanisms, fragmentation of institutions with limited or no communication between them, lack of metrics and data that curtails any evaluation of the effectiveness of policies, and lack of trust between different actors such as governments, the civil and private sectors. As a solution, this report proposes an increasingly holistic, multi-disciplinary, multi-stakeholder and agile cooperation mechanism leading to more effective outcomes.~\cite{UNSG2019}

Our proposed framework will contribute to solving the challenges of a fragmented governance landscape with isolated actors and governance tools defined and applied through often inefficient governance mechanisms. It serves as a practical instrument for identifying relevant actors and governance tools, and to coordinate these through inclusive and effective governance mechanisms applied on different levels. Three aspects ensure this: First, formal links between actors on different levels enable information flow and participation in decision-making processes. Second, soft law alternatives to existing hard law-based regulation by regulatory bodies and courts facilitates an increase in the responsibility and accountability of the private sector and of individuals. Soft law refers to standards, professional codes of conduct, insurance policies, best practices, and other mechanisms, which serve as guidelines and constraints but are not legally binding. Third, agile and dynamic governance enables managing the fast-evolving development and flexibly of measures taken in the event of unanticipated consequences after deployment of an AIS. Iterative approaches to policymaking through experiments, sandboxing and multilevel governance contribute to effective overall governance of AIS. A range of different proposals that partly include the three aspects above for the governance of AIS have been proposed or already established. However, to our knowledge, to date a framework for a comprehensive mapping of relevant actors on different decision-making levels together with mechanisms to facilitate exchange and governance tools for guidance is missing.

The remainder of this article is structured as follows: In Section~\ref{sec:GovAIS}, we elaborate on components for comprehensive technology governance including agile and adaptive governance, additional approaches such as soft law and the oversight of technological research and development through testing, compliance and review mechanisms. These components build the conceptual basis for the architecture of the multilevel governance framework for AIS outlined in Section~\ref{sec:Framework}. The framework includes various decision-making levels, relevant actors with tools that guide their behavior, possible governance mechanisms to shape the use of these tools and good practices that enable effective and efficient responses to changing conditions. Section~\ref{sec:Framework_Development} shows the application of the framework for the governance and oversight of AIS development, followed by conclusions in Section~\ref{sec:Conclusion}.\footnote{Instead of artificial intelligence (AI) we use the term AIS because intelligent systems involve at least data and algorithms for autonomous behavior. As a system is generally embedded in a real or virtual environment, additional software or hardware interfaces are necessary. Thus, the consideration of AIS includes AI as an ambiguous concept and technical components such as software, hardware and data. When referring to sources, we use the original terminology.}

% ----- %%%%%%%%%%%%%%%%%%%%%%%%%%%%%%%%%%%%%%%%%%%%%%%%%%%%%%%%%%%%% ----- % 
% ----	SECTION Governance for AIS
% ----- %%%%%%%%%%%%%%%%%%%%%%%%%%%%%%%%%%%%%%%%%%%%%%%%%%%%%%%%%%%%% ----- %

\section{Comprehensive Governance of Emerging Technologies}
\label{sec:GovAIS}

\subsection{Adaptive Governance beyond Governments}

Instead of understanding governance as a synonym for government as in the concept of the minimal state, governance under the socio-technical perspective considers the involvement of multiple actors~\cite{Rhodes1996}. This perception of governance is characterized by interdependence, shared goals and the dissolution of formal boundaries between the public, private and civil sectors featuring \enquote{new forms of action, intervention and control}~\cite{Rhodes1996}. Rhodes conceptualizes this form of governance as \enquote{self-organizing, inter-organizational networks}~\cite{Rhodes1996}. Different definitions and understandings of governance under this broader definition exist, however most of them share the notion of interdependence of relevant organizations, interactions between network stakeholders, the importance of trust for interaction and a significant degree of autonomy from the state even in self-organizing networks~\cite{Rhodes1996}.

Tucker builds on Rhodes' concept of governance as a \enquote{set of complex sociopolitical administrative interactions}~\cite{Tucker2012} and applies it to the governance of dual use technology with uncertain risks. He identifies actors in the governance network as scientists, engineers, policymakers, regulators and civil society groups that either advocate for the introduction of new technology or raise concerns about it. He highlights the importance of both anticipatory governance, which addresses potential risks of technologies at the research and development stage and adaptive governance during and after the deployment of technology. Adaptive governance includes regular technology assessment with information collection, evaluation and subsequent modification of rules.~\cite{Tucker2012}

Concepts of governance vary from state-centric governance systems over the inclusion of non-governmental actors towards the idea of governance as the management of networks. Governance systems differ in their ability to govern complex systems with \enquote{change, uncertainty, and limited predictability}~\cite{Galaz2008}. Galaz proposes a way to link a complex-systems' analysis with adaptive governance and lists three particular issues that the governance of complex and adaptive systems feature: The policy process alternates between stability and abrupt change, interactions between system components take place on multiple levels resulting in cross-scale interaction effects, and complex adaptive systems exhibit tipping points and cascading effects. To confront these challenges, Galaz proposes the idea of adaptive governance that copes with the complexity of the uncertain systems that it regulates and which relies on a balance between exploitation and exploration. While exploitation comprises mechanisms to \enquote{ensure cooperation among actors in a governance system}~\cite{Galaz2008}, exploration refers to gathering information about ongoing processes in communities that are used to \enquote{testing, evaluating, refining, and reapplying new forms of governance, institutional configurations, policies, and practices within a given policy area}~\cite{Galaz2008}. Our proposed governance framework builds on this broader definition of governance as networked actors and incorporates governance mechanisms for exploration and exploitation for effective adaptive governance.

\subsection{Agile Governance}

The introduction of agile processes enables increased adaptability of governance approaches for complex systems through flexible and dynamic governance mechanisms. In 2016, the World Economic Forum (WEF) Global Agenda Council on the Future of Software Development and Society took up the principles of the prominent \textit{Manifesto for Agile Software Development}, which are \textit{Individuals and interactions over processes and tools}, \textit{Working software over comprehensive documentation}, \textit{Customer collaboration over contract negotiation}, and \textit{Responding to change over following a plan}.~\cite{Manifesto}. The WEF applied these to the field of governance in its report \textit{A Call for Agile Governance Principles}. This report highlights the potential of agile governance principles as they \enquote{can lead to improved efficiency, public services, and public welfare, better equipping government agencies to respond to change}~\cite{WEF2016}. This would be of particular importance as governments need a framework for coping with potentially disruptive technological change, but also for efficient execution of government functions for the benefit of citizens and to enable cooperation between the government and the tech community. The WEF further highlights the need for robust, adaptable and responsive governance systems. In the style of the original manifest, the WEF-report proposes the four principles \textit{Outcomes over rules}, \textit{Responding to change over following a plan}, \textit{Participation over control}, and \textit{Self-organization over centralization}.~\cite{WEF2016}

According to the WEF, methods for agile governance would allow \enquote{a shift from planning and controlling to piloting and implementing policies to get rapid feedback and iteration}~\cite{WEF_Gov_2018}. Proposed governance mechanisms for increased agility include regulatory sandboxes as safe spaces for companies, the use of technology for more agile and distributed governance processes, the promotion of governance innovation, crowd-sourced policy-making, increased collaboration between regulators and innovators, and public representation in technology assessment. In addition, the WEF calls for an expansion of governance beyond the government including industry self-regulation, ethical standards for development, collaborative governance ecosystems and the need for transparency and trust in technology innovation.~\cite{WEF_Gov_2018}

% ----- %%%%%%%%%%%%%%%%%%%%%%%%%%%%%%%%%%%%%%%%%%%%%%%%%%%%%%%%%%%%% ----- % 
% ----	SECTION Components of Governance Frameworks
% ----- %%%%%%%%%%%%%%%%%%%%%%%%%%%%%%%%%%%%%%%%%%%%%%%%%%%%%%%%%%%%% ----- %

\subsection{Additional Components for Comprehensive Governance}
\label{sec:Components}

AIS are in general not operating in a vacuum absent any governance measures. Rather a lack of effective cooperation, oversight, and enforcement mechanisms exist. In principle, existing policies and laws within the particular domain in which AIS are used, apply to AIS as well. However, legally binding hard law is not necessarily the most effective tool for governance due to potentially undisclosed AIS use in applications, information asymmetry, and long-lasting and indirect effects. A variety of additional governance tools from soft law exist and a comprehensive governance framework needs to rely on a combination of tools to guide decisions of various actors. Besides soft law, also risk review of technologies and process governance with oversight of development are important measures to effectively guide AIS development and deployment.

\subsubsection{Soft Law Governance Tools}

Marchant defines the challenge of slow oversight mechanisms and outdated regulatory frameworks to cope with the high speed of science and technology development as the pacing problem~\cite{Marchant2011}. In order to increase the flexibility of a regulatory system to cope with challenges posed by new technology, he proposes innovations for expediting rule-making, self-regulation or cooperative regulation, issue-specific legislative initiatives, specialized courts, sunset clauses with automatically expiring legislation after a predefined period, regular reviews of programs, new independent institutions, adaptive management of policies and principles-based regulation implementing general principles rather than detailed rules. 

In addition to adjusting hard law policy-making to allow more flexibility, governance can also take other forms than legally enforceable rules. Soft law with non-legally binding expectations and tools is a powerful alternative to hard law for the governance of AIS~\cite{MarchantSL2019}. The pacing problem and the fact that AIS are often internationally developed, lead to a growing belief that \enquote{new institutions and methods that are more agile, holistic, reflexive, and inclusive}~\cite{Wallach2019} must be set up for comprehensive and effective governance of AIS. Several examples for tools of soft law such as voluntary programs, standards, codes of conduct, best practices, certification programs, and guidelines have been proposed~\cite{Wallach2019}. They exist on different levels as international rules, national principles, industry guidelines, company codes of conduct or principles for researchers and machine learning engineers such as the TechPledge~\cite{Techpledge} or the prominent Montreal Declaration for responsible AI development~\cite{Montreal2018}. A quantitative study of guidelines and soft law measures up to April 2019 identifies the five most prominent ethical values and principles as transparency, justice and fairness, non-maleficence, responsibility, and privacy~\cite{Jobin2019}. 

However, scholars highlight the importance to critically assess norm development from the private or civil sector. Key considerations are whether a process for creating guidelines is sufficiently transparent in its goals and how inclusive and open a process undertaken is~\cite{Cath2018}. Particularly given that AIS might have a strong impact on the social fabric and each individual within a society, the creation-process of design principles and ethical guidelines under which systems are evaluated, need to be transparent and balanced. Inclusivity entails a balanced age, gender and socioeconomic background of the authors. Cath highlights the importance to \enquote{remain critical of the underlying aims of AI governance solutions as well as the (unforeseen) collateral cultural impacts, particularly in terms of legitimizing private-sector-led norm development around ethics, standards and regulation} of regulatory initiatives~\cite{Cath2018}. Further critique about ethical guidelines from the private sector have raised concerns about the composition of the committees with regard to mirroring existing societal inequalities and their biases~\cite{Kane2019}. Based on an assessment of the fifteen major AI ethics guidelines in 2019, Hagendorf criticizes a lack of women’s involvement. The guidelines often being considered are imposed by institutions outside of the technical community without a clear focus and not sufficiently applying to specific situations. In addition, he calls for a stronger technological focus within AI ethics. This would better allow bridging the more abstract values towards specific recommendations for resulting requirements on the technical implementations. Further, he calls for an ethics that concentrates on individual responsibility and awareness-raising that would impact AIS development to a higher degree.~\cite{Hagendorf2019}

Wallach and Marchant~\cite{MarchantSL2019,Wallach2019} see two additional challenges with soft law. First, the lack of direct enforceability threatens the effectiveness of soft law measures. This problem, however, might be bypassed by indirect enforcement mechanisms from insurance companies, publishing institutes, grant funders or governmental enforcement programs. The second challenge of soft law is that no formal coordination between different entities that propose soft law exists. Wallach and Marchant highlight the potential of Governance Coordinating Committees for coordinating the wide range of actors in order to discuss overlaps, gaps or disagreement between the application of differing mechanisms.

\subsubsection{Risk Review of Technology}

Governance activities related to emerging and new technologies need to be based upon continuous review of the technology. This includes risk analysis of potential research findings, governance of not only the development process but also the oversight of wider consequences from the use of a particular technology. From identified and evaluated risks, governance tools can be developed for mitigating risks either through their minimization, elimination or countermeasures. In this section, we show how established risk mitigation approaches for emerging technologies in the life sciences could be applied to AIS.

The Tucker Framework was proposed in 2012 as a systematic and quantitative decision framework on governance options for emerging dual use technology, with a focus on life sciences~\cite{Tucker2012}. It consists of three consecutive steps: \textit{technology monitoring} about dual use potential of technologies from the public and private sector, \textit{technology assessment} that is split into the risk assessment of a technology and its feasibility to be misused and its potential to be governed, and thirdly \textit{technology governance} in which specific governance measures are developed according to the assessment. The decision on which measures are then proposed is based on a cost-benefit analysis in which the benefits of a technology, costs of the governance measure and stakeholder attitudes are taken into account. The process is iteratively applied and the technology is continuously monitored regarding changes or advances that make an adjustment of the governance measures necessary. Tucker’s decision framework takes into account different levels for the implementation of the governance tools. On the international level, multilateral treaties or informal coalitions of like-minded states might adopt common regulations or guidelines. Domestically, dedicated legislation could be developed and even serve as a model that proliferates to other countries. Additionally, professional associations and industry consortia could create voluntary guidelines that set targets for companies, people and products.~\cite{Tucker2012}

Risk assessment can be performed by oversight mechanisms of international organizations. For instance, the World Health Organization (WHO) set up the Expert Group on Ethics and Governance of Artificial Intelligence for Health in 2019 with the goal to make recommendations on an ethics and governance framework for oversight, development and use of artificial intelligence across health care and public health~\cite{WHOAI2019}. Besides advisory committees in international organizations, technological development can also be reviewed by national mechanisms such as ministries, councils or national advisory bodies. As an example, France established the National Advisory Council for Biosecurity that evaluates funding, execution, and dissemination of dual use research~\cite{Revill2018}. On the development level, independent review boards in companies or research institutes play an important role in responsible innovation. Once established, these bodies perform ethical risk assessments of research on innovative products and facilitate experience exchanges between developers and with decision-making bodies~\cite{Wallach2019}.

\subsubsection{Process Governance and Oversight}

Apart from dedicated bodies applying soft law governance tools and risk review of final products, the consideration of processes for the design, development and use of AIS are important, particularly for AIS whose development is distributed. Wallach and Marchant propose process-based soft law, which refers to the incorporation of ethical considerations into engineering processes and the oversight of AI development~\cite{Wallach2019}.

The research and development cycle of a technology with respect to potential oversight and governance mechanisms was mapped during the international workshop \textit{The Governance of Dual Use Research in the Life Sciences: Advancing Global Consensus on Research Oversight (2018) for exemplary use cases of life sciences}~\cite{Revill2018}. The cycle included five stages: conception and project development, funding, research, dissemination of results, and translation and product development. Governance activities guide behavior and decisions at each stage. In addition to the five stages, participants of the workshop highlighted the importance of permanent activities such as advisory boards, outreach from governments research communities, and self-governance with measures developed by the research community~\cite{Revill2018}. Potential extensions for the application of the development of AIS would involve the stages for training an AIS, and its deployment and use by a professional with less technical but rather domain expertise as in the case of doctors for medical devices.

For process oversight, several actors can be responsible. National advisory boards, funding agencies, journal publishers, export control regimes and governments certainly play an important role here~\cite{Revill2018}. However, the organizations that conduct research, development and deployment themselves are required to incorporate measures of self-governance. Linking decision-making levels can be a strong contribution towards this goal. Particularly the roles of a higher-level link to a lower level with accountability for decisions at the lower-level is proposed to ensure responsible design and use of systems. In addition, methods from agile governance such as iterative and collaborative approaches would help so that such mechanisms stay flexible and adaptable. Furthermore, organizations themselves could incorporate technical review boards as described above for ethical risk assessment. However, it is important that innovation is not hindered through process soft-governance measures~\cite{Wallach2019}.

As demonstrated in this section, different governance tools ranging from legally binding obligations and laws to ethical guidelines and voluntary commitments are available. This variety of governance tools need to be utilized for a comprehensive and adaptive governance. Thus, in the proposed governance framework in the following section, we incorporate hard and soft law governance tools together with inclusive mechanisms to create and refine the tools. In order to cope with unknown or unintended consequences and misuse potential, the framework provides an architecture to coordinate the fragmented actors for AIS governance and facilitate exchanges of experience between them. As the pacing problem increasingly challenges existing governance regimes, agile principles can be applied for higher adaptability. In addition to governance of the technology itself, mechanisms for the oversight of development processes and organizational design coupled with risk review of a technology are necessary for the comprehensive governance of AIS.

% ----- %%%%%%%%%%%%%%%%%%%%%%%%%%%%%%%%%%%%%%%%%%%%%%%%%%%%%%%%%%%%% ----- % 
% ----	SECTION The Framework
% ----- %%%%%%%%%%%%%%%%%%%%%%%%%%%%%%%%%%%%%%%%%%%%%%%%%%%%%%%%%%%%% ----- %

\section{A Practical Multilevel Governance Framework for AIS}
\label{sec:Framework}

We propose a multilevel governance framework addressing the necessity to coordinate actors on various levels to comprehensive AIS governance. It comprises an inclusive governance architecture consisting of six linked levels. The decisions of actors within the different levels are guided by tools for which the framework considers both soft and hard law measures to ensure effective governance. The framework allows the identification of relevant actors on each level, develops and evolves their decision-making tools and offers a structured approach to set up mechanisms as formalized links between the levels, enabling the governance and oversight of tools and AIS design, development, deployment and use. Additional good practices and agile approaches contribute to the efficient functioning of the mechanisms and complement the framework. In this section, the components of the framework are introduced. The subsequent Section~\ref{sec:Framework_Development} applies the framework for the development stage with a focus on the industry and standard-setting perspective.

\subsection{Levels for Decision-Making}

\begin{figure}[h]
	\centering
	\includegraphics[width=\columnwidth]{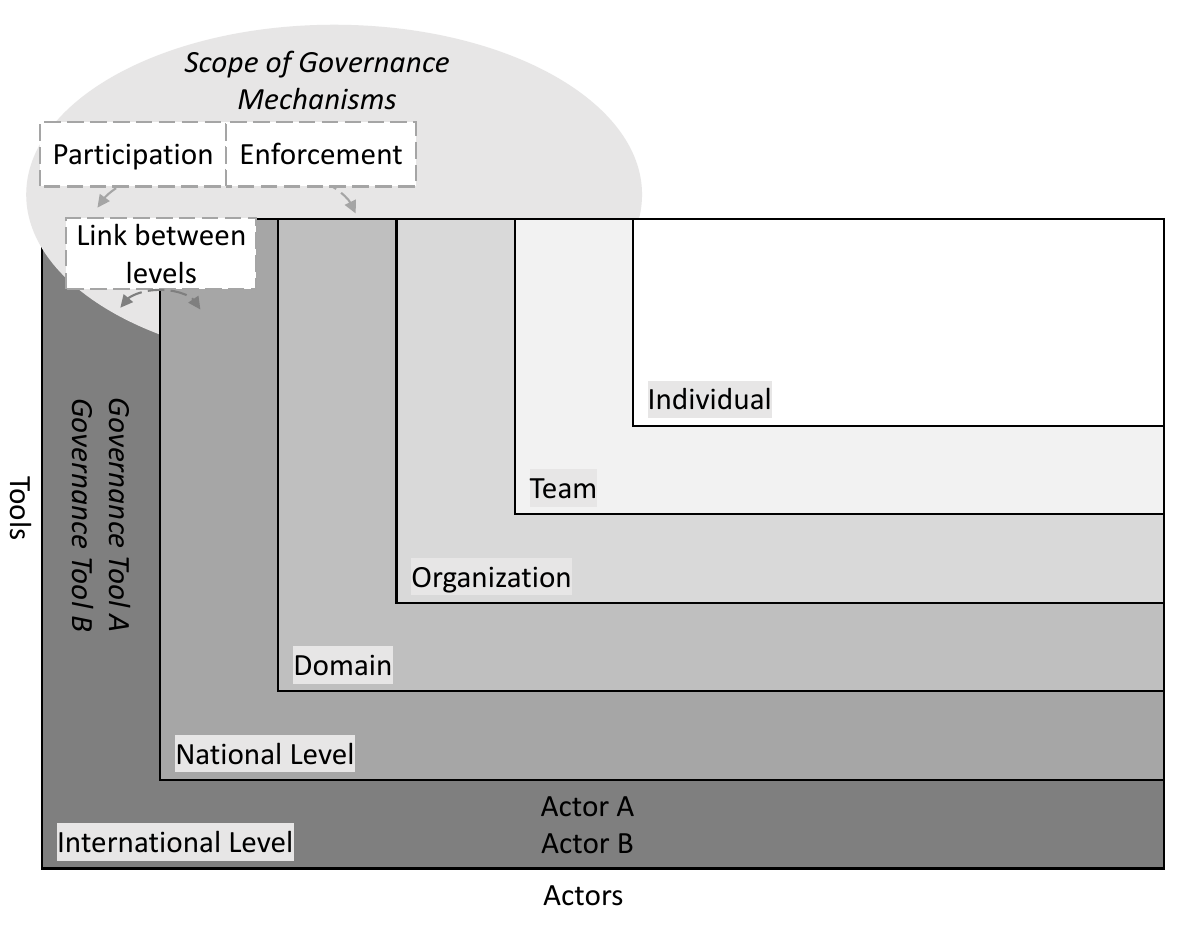}
	\caption[Framework]{Schematics of the multilevel governance framework for AIS with levels, actors, tools and mechanisms for governance and oversight.}
	\label{fig:Framework}
\end{figure}

The framework, which is depicted in Fig.~\ref{fig:Framework}, is an inclusive multilevel governance architecture with linked levels. As an architecture, it allows the identification of relevant actors on various levels together with their decision-making tools for AIS development and deployment, and the governance mechanisms that connect and facilitate interactions between the actors.

Governance tools are usually specific to particular actors. Thus, it seems reasonable to design a governance framework as a multilevel system with consideration of links between them as mechanisms. This structure allows one to identify relevant actors on different levels and their sphere of influence through governance and oversight mechanisms, such as enforcement and possibilities of their participation in higher-level decision-making processes. Actors range from international or national policy-making and standardization bodies, to industry consortia and associations companies building or deploying AIS project teams that develop, deploy or use them and individuals such as developers, end users such as doctors, marketing professionals or citizens in the innermost level of the framework.

\subsection{Tools for Guidance}

During the design, development and deployment phase of AIS, decisions are made by the actors on each level that influence the processes at each stage in the life cycle of an AIS. As described above, the decisions are guided by tools, which can either be hard law with legally binding rules or soft law such as ethical guidelines, non-legally binding expectations, company codes of conduct, formal standards or self-commitments from individuals. The effectiveness of a tool, in particular for non-legally enforceable soft law measures, depends on its publicity, applicability, usefulness, and acceptance by the relevant actors within its level.

\subsection{Mechanisms for Governance and Oversight}

The actors on the different levels in our proposed governance architecture are connected through several links. We allow that not only neighboring levels are linked but rather links and information channels bypassing a level are possible by direct participation in an outer level or within a level in the case of peer-review mechanisms. Links within the framework are governance mechanisms that facilitate the interaction of actors on different levels.

On the one hand, mechanisms allow higher level actors to exercise power onto a lower level through the establishment, refinement or enforcement of governance tools. It could also mean that oversight of a technology, processes or research is performed by a body dedicated to that task. On the other hand, linkages from lower to higher levels are participatory mechanisms that
facilitate the bottom-up information flow or enable the contribution of proposals of lower-level actors to higher decision- making bodies on a management or policy-making level. Examples here are giving feedback about a governance tool that was created from an outer level or to facilitate direct participation in the development process of a governance tool at a national or international level.

In short, governance mechanisms allow balancing  the power of the different actors regarding the creation and application of governance tools and processes across the levels. They can be grouped into spheres of influence determining their scope such as organizational design, national or international coordination as clarified in Section~\ref{sec:Framework_Development}.

\subsection{Additional Good Practices}

As proposed above, a variety of mechanisms can be applied to create, develop and enforce the application of governance tools for AIS. These governance mechanisms can be complemented by specific good practices from other areas such as the life sciences. Good practices include agile methods for governance and structured risk assessment with formalized and periodic review of a technology, such as with the Tucker framework and practices for bridging knowledge and language gaps between the actors of different levels and disciplines.

\subsubsection{Dynamic Governance and Agile Approaches}

To achieve a robust governance system in a dynamic environment with potentially disruptive changes, a good balance between the exploitation of existing tools and exploration of innovative governance tools needs to be ensured. This is particularly important for the rapid and distributed development of AIS. For exploration, iterative governance with experimentation and regulatory sandboxes can help to find novel and tangible tools for AIS governance. Agile approaches can be applied on all levels ranging from experimentation with different design principles in a company for its project teams, the refinement of company codes of conduct and industry guidelines with feedback from users and companies or, on the national level, by creating regulatory sandboxes for new technologies with consequent review and evaluation. Further, iterative governance that acknowledges lack of knowledge can promote establishing preliminary policies that can be updated at a later stage, if necessary.

Agile approaches do not necessarily need to be introduced on a large scale. Rather, dedicated pilot projects could be started within an organization such as a company, an industry, or the regional or national government. These have low entry barriers and can serve as lighthouse projects that create acceptance for the new governance mechanisms and governance tools should they be successful.

\subsubsection{Risk Assessment}

An important aspect of various governance mechanisms is the assessment of risks posed by technologies. As described in the previous section, the Tucker framework has been proposed and proven useful in the life sciences and can be applied for AIS equivalently. It can help to structure the conversation and formalize risk assessment for a number of the proposed mechanisms. Technical review boards and governments could use the Tucker framework to determine the potential of malicious use of certain technologies. As the framework requires monitoring a technology, assessing its misuse and evaluating its governability. Differing expertise is required for each of these assessments. In addition to hard law, the Tucker Framework enables relevant actors to develop packages with a variety of different governance tools including soft law.

Risk concerns might be raised by employees within a company, customers and citizens exposed to an AIS. It is important that concerns and complaints are channeled through dedicated focal points in companies or on the governmental level, and that the evaluation criteria and metrics from companies are transparent and audited. The inclusion of risk assessment into governance mechanisms could also be coupled with a foresight function as in the example of the WHO, described above. This would allow estimating the benefits and risks of technologies according to different future scenarios and time frames.

\subsubsection{Bridging Gaps}
\label{sec:BridgingGaps}

It is important to bridge the gap between higher and lower levels by making the tools of governance and processes understandable. The incorporation of formalized links between the levels, such as for developer participation in technological review boards, could enhance the positive exploitation of governance tools. This not only allows better bidirectional information flow but also increases trust through a participatory process in decision-making. Additionally, online training courses, best practice documents, and joint workshops help to bridge gaps in understanding. A prevalent challenge is finding a common language between different disciplines as terms often have divergent meanings. Thus, a shared understanding or mediation of language is essential. Clarification of concepts and terms across communities can open silos and break down language barriers. For example, accompanying the IEEE publication \textit{Ethically Aligned Design}, a glossary of AIS terms with their meanings for six different disciplines was published~\cite{IEEEGlossary2019}. The disciplines include ordinary language, computer science, economics and social sciences, engineering, philosophy and ethics, and international law and policy. Such glossaries are helpful for interdisciplinary teams as they prevent misunderstandings and can also help bridge the different governance levels.

One interesting observation is that the proposed governance framework features several characteristics of AIS technology. First, as an AIS consists of several components that are combined in an architecture through links, also the governance framework can be seen as an architecture taking into account the variety of relevant actors with their interaction. Second, AIS are often developed in a distributed and decentralized manner with shared expertise of developers and open-source collaboration. The governance framework also does not only concentrate on top-down governance, but is rather a multilevel governance framework with shared responsibility among actors on the different levels with an exchange between them. A particular focus is put on self-governance of the levels and participatory mechanisms for actors on lower-levels to contribute to higher-level decision-making. Third, as in agile development, principles are applied for the development of AIS, including several practices for increased agile and adaptable governance mechanisms. A last common characteristic is that certain AIS technologies include machine-learning (ML) algorithms that feature learning from errors. Through ML, AIS can adapt their behavior from interaction with their environment. In the proposed framework, iterative regulation and regulatory sandboxes are proposed that allow enhancing governance measures similar to the way as self-learning autonomous and intelligent systems adapt.

% ----- %%%%%%%%%%%%%%%%%%%%%%%%%%%%%%%%%%%%%%%%%%%%%%%%%%%%%%%%%%%%% ----- % 
% ----	SECTION Application of the Framework
% ----- %%%%%%%%%%%%%%%%%%%%%%%%%%%%%%%%%%%%%%%%%%%%%%%%%%%%%%%%%%%%% ----- %

\section{Application of the Framework for the Development of AIS}
\label{sec:Framework_Development}

\begin{figure*}[tb]
	\centering
	\includegraphics[width=\linewidth]{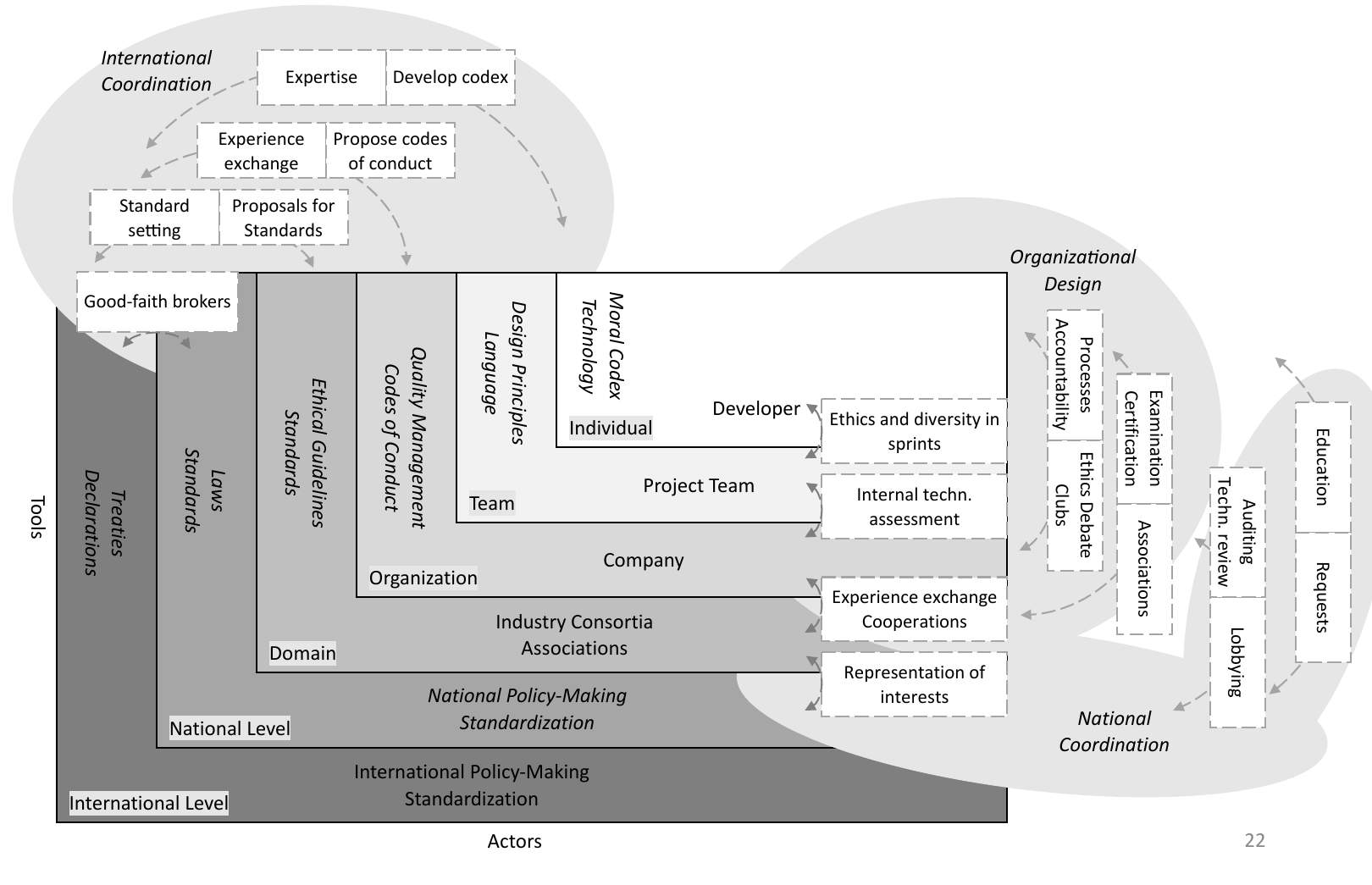}
	\caption[Framework All]{Application of the multilevel governance framework for the development of AIS with levels, actors, tools and mechanisms for governance and oversight with a focus on industries, companies and developers.}
	\label{fig:Framework_all_in}
\end{figure*}

As outlined above, the governance framework is intended to identify actors, and to develop governance tools and mechanisms for responsible design, development, deployment and use of AIS. To exemplify how the framework can be used, we apply it to the development of AIS with a focus on industry standard-setting bodies, as well as companies and developers that make up agile project teams. Fig.~\ref{fig:Framework_all_in} shows this application of the framework with prominent actors, tools and mechanisms.

\subsection{Levels for Decision-Making}
The innermost three levels of the governance framework focus on components of a company including developers, project teams and different business units. Every company is embedded in its industry, which is the fourth level. The two outermost levels represent policy-making and standardization both on the national and international level.

The actors on the individual level are developers that either contribute to the codebase, collect and pre-process data, train the AIS or design hardware components and interfaces. These individuals have varying backgrounds such as electrical or mechanical engineering, data science or computer programming. Apart from developers, an agile team generally has additional members including a product owner for strategic planning, scrum master facilitating the agile development process, designers for human-machine interaction and mechanics. Technical project teams are generally embedded in a department or business unit within a company. For simplicity, we do not focus on these levels but just consider the company as the next level
within which the project team functions. A variety of project teams and other entities such as sales, marketing, procurement, and management are part of the company-level circle.

Every company is embedded in its industry. In the example of robotics, a company is at least part of the industry circle of robotics and automation engineering. However, it also plays an important role in the domains in which the AIS is used such as the medical sector, mining or logistics. The actors on the industry level are industry consortia, associations, and other loose industry networks or affiliations. The outermost levels are policy-making and standardization both on the national and international level. They contain national policy-making bodies such as parliamentary and executive branches, and standard-setting bodies such as the Deutsches Institut für Normung (DIN) in Germany, political and economic affiliations between states such as the EU, UN or international standard-setting bodies such as the International Organization for Standardization (ISO) and the IEEE Standards Association.

\subsection{Tools for Guidance}

Governance tools guide and influence the decisions of the actors. Prominent examples are depicted at the left-hand side of each level in Fig.~\ref{fig:Framework_all_in}. It is important to note here that the diagram may create a false impression that a specific governance tool is restricted to one particular level, which is not the case. Rather, it influences the decisions of multiple actors on different levels.

Starting with the innermost level, decisions of developers are influenced by their values and moral principles, or those of the company for which they work. The formal commitment to a moral codex such as the TechPledge~\cite{Techpledge} or the Montreal Declaration for responsible AI development~\cite{Montreal2018} is both able to raise awareness and also to make implicit values explicit through specific and effective statutes. A shared ethos of AIS developers could additionally create a common understanding and awareness of responsibility within the profession. A formal pledge during education, similar to the Hippocratic Oath for medical doctors~\cite{WMA} could intensify a common sense of responsibility. In the case of software or machine learning engineers, the technical knowledge, education and available technological tools strongly influence how code is written, what the focus is during development, which considerations are taken into account and for what the code is reviewed and tested. In the case of bias in datasets or models, an increasing number of technical tools for bias checking and fairness metrics are being developed. If a developer is aware and has access to these tools and metrics, they will strongly influence the development process.

On the project team level, developers regularly engage with team members and other relevant stakeholders, such as experts in the field for which the application is being developed.  Thus, the terminology used influences the interactions of the team. In the case of diverse teams, it is important to find a common language with a shared understanding of terminology. This requires ongoing explanations and regular clarification of terms in order to avoid misunderstandings. The role of a translator that mediates between team members with different backgrounds or with outside stakeholders facilitates clarifying terminology that might be understood differently in various domains. In addition to language, the decisions of project teams are influenced by design principles. These often aim at consistency across a codebase and the seamless integration of different components within a software or hardware system. However, principles could also define mitigation approaches for challenges posed by AIS. Furthermore, reporting requirements, security and safety guidelines, and technical approaches to risk mitigation will guide the decisions of project teams.

On the organizational level, many companies have created formal codes of conducts. These influence the management in their strategic decisions, procurement teams in their sourcing-strategies, marketing on how the brand is promoted and division heads on the projects to be started. In addition to codes of conduct, codes of ethics or best practices can be defined within a company. Quality management offers a formal approach to influence product development with design principles for teams and internal review processes. Internal auditing of code, as well as final quality control for software and hardware are tools that can ensure specific quality requirements are met. Documentation of development stages such as training processes for machine learning models, applied datasets, achieved model accuracy and fairness metrics provide additional important contributions towards managing and ensuring the consistent quality of an AIS, which can be prescribed in quality management rules.

A company is embedded in its relevant industry together with its competitors and customers. Thus, from a competitive perspective, decisions of companies in an industry are influenced by peer pressure, user demand and competitive products and services. Industry standards ensure that a company fulfills certain quality and safety requirements, and therefore influence company decisions.. Another set of tools that influence company decisions can be ethical guidelines that are created and promoted by industry consortia. A prominent consortia in the field of AIS is the Partnership on AI, with more than 90 partners, which develops AI ethics guidelines as best practices for its members. Another example is the US Association for Computing Machinery that published its seven Principles for Algorithmic Transparency and Accountability in 2017~\cite{ACM2017}.

On the two outermost levels, national governments pass laws and institute regulations. International organizations facilitate the enactment of international treaties. Additionally, accredited certification bodies can certify that a company complies with certain standards. International standards can either be used directly by companies or after they have been harmonized with national standards. As an example, for the field of robotics, on of the first norms relevant to AIS outside the delimited space of industrial robotics was ISO~13482 (\textit{Robots and robotic devices - Safety requirements for personal care robots}), published in February 2014. It was harmonized in national standards as, for example, in the German norm DIN~EN~ISO~13482:2014. 

International standards related to AIS are not only directed at safety and security. The ongoing project of the standard series IEEE P7000, undertaken by the IEEE Standards Association, aims at addressing the intersection of technical and ethical considerations of AIS across domains and throughout a systems lifecycle. Relevant for the national level are international treaties such as arms control regimes that must be harmonized to national law and laws from multinational organizations, such as the GDPR of the EU that its member states and those that want to do business with the EU are required to respect. National export controls can be relevant for limiting dissemination of dual-use technologies. Needless to say, national law is relevant to all companies within a country and defines how the actors deploy and use a technology in a particular domain.

To conclude, the actors on the different levels of the governance framework have and are making important decisions on how AIS are designed, developed, deployed and used. Decisions of the actors are influenced by certain governance tools of which prominent ones are depicted in Fig.~\ref{fig:Framework_all_in}. The effectiveness of a tool, in particular in the case of soft law that is not legally enforceable, depends on its applicability, usefulness, and acceptance by the relevant actors within the respective level.

\subsection{Mechanisms for Governance and Oversight}

Fig.~\ref{fig:Framework_all_in} shows several proposed governance mechanisms with the tools they impact and the actors that take part in the mechanisms. As it can be seen, not only neighboring levels are linked but participation and information flow happen over various levels as for example individuals can engage in and contribute to decision-making on the national or international level. Mechanisms can be grouped into the three spheres of influence Organizational design, National coordination and International coordination, which are outlined in the following in detail. The proposed mechanisms are not an exhaustive list of measures for the governance of AIS development. They are rather exemplary mechanisms, which seem particularly promising and show how the framework can be used to create these mechanisms by coordinating relevant actors.

\subsubsection{Organizational Design}

Beginning at the core of the framework, developers formulate commands and write source code that determines how the AIS work. Thus, special focus needs to be put on mechanisms that shape the tools on which developers base their decisions. As described above, developers and engineers are influenced by the technology and by their personal value system, but they are at the same time embedded in the bigger circle of a project team. In the case of agile software development, development sprints are scheduled for one to four weeks after which a sprint review is held. In this review, besides the development team, a scrum master and a product owner take part. Two mechanisms could be introduced here in order to incorporate perspectives on the wider implications of AIS. One possibility is that the participants of the sprint reviews are expanded with social scientists or ethicists. These would bring in ethical and societal considerations to the otherwise rather technical and business-oriented discussion during the sprint. Another possibility is to dedicate a part of the sprint towards a critical assessment of the wider societal impacts of technical features. By bringing ethical topics and diversity of members into development sprints, both the design principles of a project team and also its language can be changed. From raised concerns, developers might also realize a need for a dedicated technical solution, for example, testing a dataset on potential biases. Finally, by reflecting on one’s responsibility and alignment to ethical guidelines or company codes of conduct, the personal values of a developer are highlighted. In case a raised concern is not acknowledged by the other members of the project team or its solution needs to be addressed on a higher level, such as on the company, industry or national level, the establishment of mechanisms for concern raising within a company such as an ombudsperson in charge of ethical concerns or channels for claims. Requests on clarification of law and the formation of industry associations on the national or international level are all potential means for bottom-up information flow.

Within a company, governance tools influence the development stages of products, starting with the creation and approval of business ideas and including requirements for the design of an AIS and how they are finally developed and tested through dedicated procedures. The first group of mechanisms that strongly governs how development is executed within a company are its internal processes. It makes a difference, whether a developer is being held responsible for the code he or she writes, or rather considers him- or herself as only a small gear in the machinery for whom as soon as the code is integrated into the bigger codebase, he or she looses any responsibility. If responsible development is the aim of a company, a certain level of ownership should be established for a developer and a project team. However, as one developer might only implement a limited number of aspects and components of a bigger product, a dedicated unit within a company is important for a comprehensive risk review of a product. A technology review board could review research proposals and product ideas before the development stage and continue with audits of new features with respect to challenges and risks in a pre-deployment phase. However, all risks and consequences cannot be anticipated and mitigated. Thus, it is particularly important to monitor AIS products after deployment regarding its impact on individuals and communities, its social acceptance and any public concerns raised. Feedback from customers’ experience and a continuous review after deployment allows a company to improve the system design and tailor future quality requirements. With suitable processes, these learnings would then propagate down to project teams in the form of adjusted design principles and quality standards. Members of a company technology review board could engage on the level of project teams during a sprint review and raise ethical, societal impact or risk concerns about new features. Conversely, developers could also delegate technical experts as representatives to the technology review board, in order to bring technical competence from project work to the board.

Besides a formalized review board, informal groups could be created within companies that exchange their experience and discuss ethical and risk concerns about the products. Through such platforms, team members could be encouraged to debate current or potential future ethical dilemmas together with ethicists or social scientists. The basis for responsible development, however, is that an employer grants explicit resources and dedicated time to employees to reflect, and more importantly, to take measures towards responsible technological development that requires more effort than achieving the bare functional requirements. The continuous monitoring of processes, transparency measures, ethical debate clubs, and resource allocation for responsible development might be considered as a burden by a company. However, it could give a company also higher credibility and an improved reputation that results in a competitive advantage towards other companies in the same industry, as well as help to prevent later customer criticism or liability risks.

An important governance mechanism that influences the strategies of companies is exposure to competitors in the same industry. Exposure can have various forms such as deliberate experience exchange, competition, the formation of associations and industry consortia, or the joint development of standards. Deliberate experience exchange on safe and responsible AIS development could be promoted between companies in the same industry. For example, individual company codes of conduct could be aligned for the creation of industry guidelines as in the case of the Partnership on AI initiative. This would counteract the increasing fragmentation and multiplication of ethical guidelines from industries, companies, and NGOs with differing codes of conduct. It is important that a consolidation of guidelines includes a broad range of actors and is developed through transparent processes. The facilitation by a respected organization, such as in the chemical sector with the Hague Ethical Guidelines created under the umbrella of the Organization for the Prohibition of Chemical Weapons (OPCW)~\cite{OPCW2019}, can be a helpful mechanism for broad acceptance and transparency. Additionally, best practices on quality management or process monitoring could be exchanged within industry consortia. A dedicated group might be created which brings together representatives from several company technology review boards in a new industry-wide exchange. Such a group could facilitate experience exchange and early-detection of unintended consequences. Additionally, technical problems and societal challenges after deployment could be discussed within this group and be fed back to the companies’ project teams.

\subsubsection{National Coordination}

Intensified cooperation and experience exchange between companies could also result in the creation of national standards, or to identify common challenges shared by companies of a certain industry. This could be the identification of necessary legislation to ensure a level playing field. It might also be the incentivization of behavior away from immediate competitive advantage by helping companies within an industry in a collective way. Such interests can be brought forward on a national level either through lobbying of individual companies or through industry interest groups. An example of this would be to set strict rules for the utilization of personal medical data with high security requirements such as for cancer detection in medical images through classification models. For a particular company, high requirements will result in increased effort and costs for an application. However, if all companies are legally obliged to follow certain standards on data security, for example, the burden is evenly shared among all competitors without giving a disadvantage to those companies setting high-quality standards on data protection. This would make potential data breaches less likely and in the medical industry build trust between patients to share data. Helping companies that require data for training accurate models and benefiting the industry as a whole. For research-focused companies, voluntary peer-review of development processes could be established within a research network. Such a network could also take on an important role in the certification of companies and products providing a competitive advantage.

Education and awareness-raising are important mechanisms on the individual level. If, for example, a developer is sensitized about a potential societal impact of his or her actions and technological products, then it is more likely that he or she will naturally reflect on their contribution and raise concerns within a project team. Any meaningful participation, such as an active engagement in the ethical part of a sprint, requires that the team members have a certain level of awareness of the wider impact of AIS. Early education in schools and particularly at the university level in technical courses is important. Ethics courses for technologists working in an industry need to start at an early stage in their career. Early reflection and opinion formation with peers during education can motivate developers to raise concerns with a high degree of commitment throughout their career.

To close the gap of the pacing problem, a certain amount of technical expertise and advice for policy-makers is essential. Informed advice can help to create solid scientifically-based regulation. However, every actor brings in their own interests and subjective science-based advocacy could lead to biased advice and policy-making. Thus, a balanced view and diverse representation from different industries, key companies, civil society and NGOs must be ensured as well as input from varying gender and socioeconomic backgrounds.

Governments need to build up own competence and expertise within ministries and throughout agencies. A national technical review body for AIS would facilitate the monitoring of AIS after deployment regarding unforeseen, unintended or neglected consequences. Such a national body could also be a focal point for citizen complaints or inquiries regarding adherence to standards and compliance to laws by certain technologies. Government agencies and the oversight body would additionally undertake regular outreach activities to the research and development communities for experience exchange and to fulfill auditing activities. Apart from oversight and policy-making, governments can also shape industry standards and foster adherence to them. In the case of contracting, a government can set particular contractual requirements. This can lead to the development or modification of industry standards. Additionally, terms and conditions for the funding of research activities can require risk mitigation strategies and adherence to guidelines from industry and standard-setting bodies such as ISO and IEEE. Governments also need to actively promote research on technical measures for risk mitigation of AIS through dedicated grants.

\subsubsection{International Coordination}

The third sphere of influence for governance mechanisms is the international level. This does not only comprise the coordination of governments but also international standard-setting bodies, institutes for certification and international organizations. Scientific or industry advice is also relevant in international discussions and conferences. In addition, review of many technologies and the oversight of development need to take place on the international level as multinational corporations operate on a global level across borders, with products and AIS often being rolled out globally with only minor differences and adjustments in hardware and software.

Supranational fora such as the EU Initiative on AI can be used as a governance mechanism to mediate between different levels and thus facilitate information flow from lower-levels such as developers, companies, industry groups and governments to the international level. This can be achieved either by seminars and workshops with the different groups, international committees that aim at representing the diverse input of lower-level actors such as in the EU High-Level Expert Group on AI, or by collecting feedback for policy-proposals through online collaboration tools.

A risk management body similar to the WHO Expert Group on Ethics and Governance of Artificial Intelligence for Health could help in establishing interoperability of systems, setting standards and proposing international rules. However, a outstanding question is whether an international treaty around AIS with a dedicated organization, such as the OPCW, would be supported by a large number of states. Clearly, such a body could help in the creation of prominent and consolidated ethical guidelines. However, it is questionable if the international community would be able to negotiate a common understanding and scope for required regulation and norms of AIS. In any case, the international community should leverage the existing ecosystem including the International Telecommunication Union and other relevant UN bodies such as the United Nations Office for Disarmament Affairs (UNODA) to define responsibilities on particular aspects of AIS governance. This is essential to avoid further fragmentation of international governance tools and actors.

The establishment of international networks, such as the professional organization IEEE, and conferences, like the AI for Good Global Summit, allows pooling the input of a wide group of different actors on the global level. Within these international conferences and organizations, norms and challenges can be discussed under different perspectives and cultural backgrounds. Low entry barriers need to be ensured to allow the participation of diverse actors, and an honest debate on AIS chances and challenges. Through these activities, governance tools can be created in the form of statutes, pledges, formal standards or declarations, commitments by industries to self-governance, and by governments to established regulations and regulatory agencies. Globally, an international governance coordination committee such as the International Congress for the Governance of AI~\cite{Wallach2019}, can be able to mediate between different actors as a good faith broker. A diverse and inclusive composition of participants is crucial for efficient monitoring technical developments and the creation of respected and effective governance.

% ----- %%%%%%%%%%%%%%%%%%%%%%%%%%%%%%%%%%%%%%%%%%%%%%%%%%%%%%%%%%%%% ----- % 
% ----	SECTION Conclusion
% ----- %%%%%%%%%%%%%%%%%%%%%%%%%%%%%%%%%%%%%%%%%%%%%%%%%%%%%%%%%%%%% ----- %

\section{Conclusion}
\label{sec:Conclusion}

AIS are increasingly applied in a wide range of different domains, such as the medical sector, and soon they will be ubiquitous as personal assistants in the home. These applications offer considerable benefits. However, several characteristics of AIS such as potential unintended large-scale consequences, potential weaponization or malicious misuse, and the unpredictability of self-learning systems pose challenges for the current governance ecosystem. Weak institutions, low technical expertise, together with slow regulatory processes by policy-makers are inappropriate for the effective governance of rapid and distributed AIS development. At the same time, an increased number of initiatives and proposals for new forms of governance and industry self-regulation together with a multilevel suite of governance tools offer promise for guiding decisions in AIS research and development. 

As a practical contribution towards mapping and coordinating the global governance landscape towards effective AIS governance, we propose a multilevel governance framework. It can facilitate the identification of relevant actors and the development of tools that guide their decisions and behavior. Furthermore, we propose governance mechanisms that allow formal links and interaction between the actors. Mechanisms allow higher levels to propose or enforce the application of governance tools or to establish oversight processes for actors on a lower level. Governance mechanisms also serve as links from lower levels to higher levels such as participatory mechanisms that facilitate bottom-up information flow or enable the contribution of proposals of individuals to higher decision-making bodies. The governance framework is complemented with good practices and elements from agile governance such as iterative policymaking, regulatory sandboxes, and periodic reviews of governance tools. These practices both contribute to the efficient functioning of the proposed governance mechanisms and facilitate adaptive governance.

The proposed governance framework outlined is intended to be a practical contribution towards the responsible design, development, deployment, and use of AIS. It allows the identification of relevant actors, develops and evolves their decision-making tools and sets up mechanisms for the governance and oversight of these tools and technology. Additionally, good practices contribute to the efficient functioning of the mechanisms. We encourage actors on the different levels including developers, private companies, governments and international organizations to apply this governance framework to ensure responsible, beneficial and sustainable application of AIS.

\printbibliography

@techreport{WEF2018,
address = {Geneva},
author = {Herweijer, Celine and Combes, Benjamin and Ramchandani, Pia and Sidhu, Jasnam},
institution = {World Economic Forum},
issn = {20754124},
title = {{Harnessing Artificial Intelligence for the Earth}},
url = {http://www3.weforum.org/docs/Harnessing_Artificial_Intelligence_for_the_Earth_report_2018.pdf},
year = {2018}
}

@techreport{Chu2018good,
author = {Chui, Michael and Harryson, Martin and Manyika, James and Roberts, Roger and Chung, Rita and van Heteren, Ashley and Nel, Pieter},
publisher = {McKinsey Global Institute},
institution = {MGI},
title = {{Notes from the AI Frontier - applying AI for Social Good}},
year = {2018}
}

@techreport{Brundage2018,
archivePrefix = {arXiv},
arxivId = {1802.07228},
author = {Brundage, Miles and Avin, Shahar and Clark, Jack and Toner, Helen and Eckersley, Peter and Garfinkel, Ben and Dafoe, Allan and Scharre, Paul and Zeitzoff, Thomas and Filar, Bobby and Anderson, Hyrum and Roff, Heather and Allen, Gregory C and Steinhardt, Jacob and Flynn, Carrick and H{\'{E}}igeartaigh, Se{\'{a}}n {\'{O}} and Beard, Simon and Belfield, Haydn and Farquhar, Sebastian and Lyle, Clare and Crootof, Rebecca and Evans, Owain and Page, Michael and Bryson, Joanna and Yampolskiy, Roman and Amodei, Dario},
eprint = {1802.07228},
month = {2},
title = {{The Malicious Use of Artificial Intelligence: Forecasting, Prevention, and Mitigation}},
url = {http://arxiv.org/abs/1802.07228},
year = {2018}
}

@article{Gao2018,
archivePrefix = {arXiv},
arxivId = {arXiv:1802.06840v1},
author = {Gao, Yang and Singh, Rita and Raj, Bhiksha},
doi = {10.1109/ICASSP.2018.8462018},
eprint = {arXiv:1802.06840v1},
isbn = {9781538646588},
issn = {15206149},
journal = {ICASSP, IEEE International Conference on Acoustics, Speech and Signal Processing - Proceedings},
pages = {2506--2510},
title = {{Voice Impersonation Using Generative Adversarial Networks}},
volume = {2018-April},
year = {2018}
}

@misc{POEHLER2018GGE,
title={A Technological Perspective on Misuse of Available AI}, 
author={Lukas P\"ohler and Valentin Schrader and Alexander Ladwein and Florian von Keller},
booktitle = {UN Meeting of the Group of Governmental Experts on LAWS},
year={2018},
doi = {10.48550/arXiv.2403.15325},
eprint={2403.15325},
archivePrefix={arXiv},
primaryClass={cs.CY}
}

@article{Kim2018Deep,
author = {Kim, Hyeongwoo and Garrido, Pablo and Tewari, Ayush and Xu, Weipeng and Thies, Justus and Nie{\ss}ner, Matthias and P{\'{e}}rez, Patrick and Richardt, Christian and Zollh{\"{o}}fer, Michael and Theobalt, Christian},
doi = {10.1145/3197517.3201283},
issn = {07300301},
journal = {ACM Transactions on Graphics},
month = {5},
number = {4},
title = {{Deep Video Portraits}},
url = {http://arxiv.org/abs/1805.11714},
volume = {37},
year = {2018}
}

@techreport{Pringle2017,
author = {Pringle, Ramona},
booktitle = {IEEE Society on Social Implications of Technology},
title = {{Unintended Consequences of Living with AI}},
institution = {IEEE},
url = {https://technologyandsociety.org/unintended-consequences-of-living-with-ai-the-paradox-of-technological-potentialpart-ii/},
year = {2017}
}

@book{Brockmann2019,
address = {Solna},
author = {Brockmann, Kolja and Bauer, Sibylle and Boulanin, Vincent},
publisher = {Stockholm International Peace Research Institute},
title = {{BIO PLUS X Arms Control and the Convergence of Biology and Emerging Technologies}},
url = {https://www.sipri.org/publications/2019/other-publications/bio-plus-x-arms-control-and-convergence-biology-and-emerging-technologies},
year = {2019}
}

@book{Sipri2019,
address = {Solna},
edition = {May 2019},
editor = {Boulanin, Vincent},
publisher = {Stockholm International Peace Research Institute},
title = {{The Impact of Artificial Intelligence on Strategic Stability and Nuclear Risk}},
url = {https://www.sipri.org/sites/default/files/2019-05/sipri1905-ai-strategic-stability-nuclear-risk.pdf},
volume = {I},
year = {2019}
}

@techreport{IEEEGlossary2019,
author = {Mattingly-Jordan, Sara and Day, Rosalie and Donaldson, Bob and Gray, Phillip and Ingram, L. Maria},
publisher = {IEEE},
institution = {IEEE},
title = {{Ethically Aligned Design First Edition Glossary}},
year = {2019}
}

@techreport{Cummings2017,
address = {London},
author = {Cummings, Mary L.},
booktitle = {International Security Department and US and the Americas Programme},
doi = {10.1145/2046684.2046699},
institution = {Chatham House},
isbn = {9781450310031},
title = {{Artificial Intelligence and the Future of Warfare}},
url = {http://dl.acm.org/citation.cfm?doid=2046684.2046699},
year = {2017}
}

@article{Gill2019,
author = {Gill, Amandeep Singh},
doi = {10.1017/S0892679419000145},
issn = {17477093},
journal = {Ethics and International Affairs},
number = {2},
pages = {169--179},
title = {{Artificial Intelligence and International Security: The Long View}},
volume = {33},
year = {2019}
}

@techreport{Rickly2017,
address = {Geneva},
author = {Rickli, Jean-Marc},
institution = {Geneva Centre for Security Policy (GCSP)},
title = {{Defence Future Technologies What we see on the horizon}},
year = {2017}
}

@techreport{Hagendorf2019,
archivePrefix = {arXiv},
arxivId = {1903.03425},
author = {Hagendorff, Thilo},
institution = {University of Tuebingen},
eprint = {1903.03425},
title = {{The Ethics of AI Ethics -- An Evaluation of Guidelines}},
url = {http://arxiv.org/abs/1903.03425},
year = {2019}
}

@article{Jobin2019,
author = {Jobin, Anna and Ienca, Marcello and Vayena, Effy},
doi = {10.1038/s42256-019-0088-2},
issn = {2522-5839},
journal = {Nature Machine Intelligence},
number = {9},
pages = {389--399},
title = {{The Global Landscape of AI Ethics Guidelines}},
url = {http://www.nature.com/articles/s42256-019-0088-2},
volume = {1},
year = {2019}
}

@book{Marchant2011,
address = {Dordrecht},
author =  {Marchant, Gary E. and Allenby, Braden R. and Herkert, Joseph R.},
doi = {10.1007/978-94-007-1356-7},
isbn = {978-94-007-1355-0},
pages = {19--33},
publisher = {Springer Netherlands},
series = {The International Library of Ethics, Law and Technology},
title = {{The Growing Gap Between Emerging Technologies and Legal-Ethical Oversight}},
url = {http://link.springer.com/10.1007/978-94-007-1356-7},
volume = {7},
year = {2011}
}

@article{MarchantSL2019,
author = {Marchant, Gary},
journal = {UCLA: The Program on Understanding Law, Science, and Evidence (PULSE)},
title = {{Soft Law Governance of Artificial Intelligence}},
url = {https://escholarship.org/uc/item/0jq252ks},
year = {2019}
}

@article{Wallach2019,
author = {Wallach, Wendell and Marchant, Gary},
doi = {10.1109/JPROC.2019.2899422},
issn = {00189219},
journal = {Proceedings of the IEEE},
number = {3},
pages = {505--508},
publisher = {IEEE},
title = {{Toward the Agile and Comprehensive International Governance of AI and Robotics}},
volume = {107},
year = {2019}
}

@article{Cath2018,
author = {Cath, Corinne},
doi = {10.1098/rsta.2018.0080},
issn = {1364503X},
journal = {Philosophical Transactions of the Royal Society A: Mathematical, Physical and Engineering Sciences},
number = {2133},
title = {{Governing Artificial Intelligence: Ethical, Legal and Technical Opportunities and Challenges}},
volume = {376},
year = {2018}
}

@techreport{UNSG2019,
author = {{UN Secretary-General's High-level Panel on Digital Cooperation}},
institution = {United Nations},
title = {{The Age of Digital Interdependence}},
url = {https://digitalcooperation.org/wp-content/uploads/2019/06/DigitalCooperation-report-for-web.pdf},
year = {2019}
}

@techreport{Kane2019,
address = {Vienna},
author = {Kane, Angela},
institution = {Vienna Center for Disarmament and Non-Proliferation},
title = {{Robotics, AI, and Humanity: Science, Ethics, and Policy}},
year = {2019}
}

@misc{Montreal2018,
address = {Montreal},
author = {{The Forum on the Socially Responsible Development of AI}},
pages = {1--21},
publisher = {Universit{\'{e}} de Montr{\'{e}}al},
title = {{Montreal Declaration for a Responsible Development of Artificial Intelligence}},
url = {https://www.montrealdeclaration-responsibleai.com/reports-of-montreal-declaration},
year = {2018}
}

@misc{Techpledge,
author = {{Techfestival 150 Think Tank}},
title = {{The TechPledge}},
url = {https://www.techpledge.org},
year = {2019}
}

@book{Scharre2018,
author = {Scharre, Paul},
title = {Army of None: Autonomous Weapons and the Future of War},
year = {2018},
isbn = {1541469682},
publisher = {W.W. Norton {\&} Company}
}

@book{Singer2009,
title={Wired for War: The Robotics Revolution and Conflict in the 21st Century},
year={2009},
isbn={9780143116844},
author={Singer, P.W.},
publisher={Penguin Publishing Group}
}

@misc{OPCW2019,
author = {{Organisation for the Prohibition of Chemical Weapons}},
title = {{The Hague Ethical Guidelines}},
url = {https://www.opcw.org/hague-ethical-guidelines},
year = {2019}
}

@book{Revill2018,
author = {{National Academies of Sciences, Engineering, and Medicine}},
address = {Washington, D.C.},
doi = {10.17226/25154},
isbn = {978-0-309-47799-4},
month = {11},
publisher = {National Academies Press},
title = {{Governance of Dual Use Research in the Life Sciences: Advancing Global Consensus on Research Oversight: Proceedings of a Workshop}},
url = {https://www.nap.edu/catalog/25154},
year = {2018}
}

@misc{WHOAI2019,
author = {{World Health Organization}},
title = {{WHO Expert Group on Ethics and Governance of Artificial Intelligence for Health}},
url = {https://www.who.int/groups/who-expert-group-on-ethics-and-governance-of-artificial-intelligence-for-health/about},
year = {2019}
}

@techreport{WEF_Gov_2018,
address = {Geneva},
author = {{World Economic Forum}},
institution = {WEF},
title = {{Agile Governance Reimagining Policy-making in the Fourth Industrial Revolution}},
url = {http://www3.weforum.org/docs/WEF_Agile_Governance_Reimagining_Policy-making_4IR_report.pdf},
year = {2018}
}

@book{Tucker2012,
address = {Cambridge},
editor = {Tucker, Jonathan B.},
isbn = {978-0-262-01717-6},
publisher = {The MIT Press},
title = {{Innovation, Dual Use, and Security: Managing the Risks of Emerging Biological and Chemical Technologies}},
year = {2012}
}

@article{Rhodes1996,
author = {Rhodes, R. A. W.},
doi = {10.1111/j.1467-9248.1996.tb01747.x},
issn = {0032-3217},
journal = {Political Studies},
number = {4},
pages = {652--667},
title = {{The New Governance: Governing without Government}},
url = {http://journals.sagepub.com/doi/10.1111/j.1467-9248.1996.tb01747.x},
volume = {44},
year = {1996}
}

@article{Galaz2008,
author = {Galaz, Victor and Duit, Andreas},
isbn = {0952-1895},
journal = {Governance: An International Journal of Policy, Administration, and Institutions},
number = {3},
pages = {311--335},
title = {{Governance and Complexity: Emerging Issues for Governance Theory}},
volume = {21},
year = {2008}
}

@misc{Manifesto,
author = {{The Agile Alliance}},
title = {{Manifesto for Agile Software Development}},
url = {http://agilemanifesto.org},
year = {2001}
}

@techreport{WEF2016,
author = {{WEF Global Agenda Council on the Future of Software Development and Society}},
institution = {WEF},
title = {{A Call for Agile Governance Principles}},
url = {http://www3.weforum.org/docs/IP/2016/ICT/Agile_Governance_Summary.pdf},
year = {2016}
}

@techreport{ACM2017,
address = {Washington, D.C.},
author = {{ACM US Public Policy Council}},
institution = {ACM},
title = {{Statement on Algorithmic Transparency and Accountability}},
url = {https://www.acm.org/binaries/content/assets/public-policy/2017_usacm_statement_algorithms.pdf},
year = {2017}
}

@misc{WMA,
author = {World Medical Association},
title = {{WMA Declaration of Geneva}},
url = {https://www.wma.net/policies-post/wma-declaration-of-geneva/},
year = {2018} 
}

@misc{Radford2019LanguageMA,
  title={Language Models are Unsupervised Multitask Learners},
  author={Alec Radford and Jeffrey Wu and Rewon Child and David Luan and Dario Amodei and Ilya Sutskever},
  year={2019}
}

\end{document}